\documentclass[11pt]{article}
\usepackage[dvips]{color}
\usepackage{epsfig}
\usepackage{amsmath}
\usepackage{graphicx}
\usepackage{bm}
\date{}
\begin{document}
\title{{\bf Bouncing scalar field cosmology in the polymeric minisuperspace picture}}
\author{B. Vakili$^1$\thanks{%
e-mail: b-vakili@iauc.ac.ir}\hspace{1.2mm}, K. Nozari$^2$\thanks{%
e-mail: knozari@umz.ac.ir}\hspace{1.2mm}, V. Hosseinzadeh$^2$\thanks{%
e-mail: v.hosseinzadeh@stu.umz.ac.ir}\hspace{1.2mm} and M. A. Gorji$^2$\thanks{%
e-mail: m.gorji@stu.umz.ac.ir}
%EndAName
\\\\
$^1${\small {\it Department of Physics, Chalous Branch, I.A.U., P.O.
Box 46615-397, Chalous, Iran}}\\$^2${\small {\it Department of
Physics, Faculty of Basic Sciences, University of Mazandaran,}}\\
{\small {\it P.O. Box 47416-95447 Babolsar, Iran}}} \maketitle

\begin{abstract}
We study a cosmological setup consisting of a FRW metric as the
background geometry with a massless scalar field in the framework of
classical polymerization of a given dynamical system. To do this, we
first introduce the polymeric representation of the quantum
operators. We then extend the corresponding process to reach a
transformation which maps any classical variable to its polymeric
counterpart. It is shown that such a formalism has also an analogue
in terms of the symplectic structure, i.e., instead of applying
polymerization to the classical Hamiltonian to arrive its polymeric
form, one can use a new set of variables in terms of which
Hamiltonian retains its form but now the corresponding symplectic
structure gets a new deformed functional form. We show that these
two methods are equivalent and by applying of them to the scalar
field FRW cosmology see that the resulting scale factor exhibits a
bouncing behavior from a contraction phase to an expanding era.
Since the replacing of the big bang singularity by a bouncing
behavior is one of the most important predictions of the quantum
cosmological theories, we may claim that our polymerized classical
model brings with itself some signals from quantum theory.

\vspace{5mm}\noindent\\
PACS numbers: \vspace{0.8mm}04.60.Nc, 04.60.Kz \newline Keywords:
Scalar field cosmology, Classical polymerization
\end{abstract}

\section{Introduction}
In the regimes that gravitational systems behave classically,
general theory of relativity presents an almost complete description
of such phenomena. Black hole physics and cosmology are perhaps the
best examples which show the power of general relativity to explain
how a system will be evolved under the act of gravity. However, it
is well-known that there are also some aspects of these systems that
general relativity fails to describe them. For example, we can refer
to the problem of black hole radiation and all types of classical
singularities that appear in the cosmological solutions to the
Einstein field equations. Indeed, a quantum theory of cosmology is
needed to understand the behavior of the universe in the vicinity of
a classical singularity and also, the black hole radiation formalism
is based on applying quantum field theory to the curved space-time
around a black hole. Therefore, any hope of dealing with such issues
would be in the development of a quantum theory of gravity which
seems to be the main challenging task of theoretical physics and a
wide variety of approaches are offered to the issue, from the DeWitt
traditional canonical approach \cite{DeWitt} to the more modern
viewpoints of string theory and loop quantum gravity
\cite{Loop}-\cite{Rov} in terms of which the space-time has a
granular structure. The granularity property of the underlying
space, in turn, supports the idea of existence of a minimal
measurable length \cite{String}-\cite{Gar}, one of the main features
of which is to deform the algebraic structure of ordinary quantum
mechanics.

One of the models that uses the idea of the existence of a minimal
measurable length scale in its formalism is the so-called polymer
quantization \cite{Corichi1}. In this approach to quantization of a
dynamical system one uses methods very similar to the effective
models of loop quantum gravity \cite{Ashtekar}. In this scenario the
main role is played by the polymer length scale in such a way that,
unlike the deformed algebraic structure usually coming from the
noncommutative phase-space variables, it enters into the Hamiltonian
of the system to deform its functional form into a new one called
the polymeric Hamiltonian. In the polymer representation of a
quantum system the deformation parameter $\lambda$ in the
Hamiltonian is responsible for the granularity property of the
underlying space. The presence of this parameter is such that in the
continuum limit when $\lambda \rightarrow 0$ the discrete geometry
becomes continuum and the physical system behaves classically.
Regards to this $\lambda$-dependent Hamiltonian, in the
corresponding Hilbert space, the momentum operator is not defined by
its normal way as in the usual quantum mechanics, instead it can be
defined by a polymerization process in which the polymer scale comes
into the play as a first-order parameter to define the derivative
operator \cite{Corichi1}, \cite{Campiglia}. With this motivations,
polymer quantization has attracted some attentions in recent years
in the fields which have to deal with the quantum gravitational
effects in a physical system. Specially, in the domain of quantum
features of cosmological models, it is shown that the
$\lambda$-dependent classical cosmology leads to a modified Friedman
equation which is very similar to the one that comes from loop
quantization of the model \cite{Campiglia}-\cite{Kar}.

Before going any further, some remarks are in order to clarify how
the polymerization process, which originally is a quantum pattern,
may work in a classical scheme. To see this, suppose a classical
dynamical system is described by the Hamiltonian $H$. In the
ordinary quantum version of this system the Hamiltonian operator
contains the parameter $\hbar$ such that in the limit $\hbar
\rightarrow 0$ the $\hbar$-dependent Hamiltonian $H_{\hbar}$ turns
to its classical counterpart $H$. This is also true for the
equations of motion, i.e., the classical equations of motion for the
mentioned system are nothing but the $\hbar \rightarrow 0$ limit of
the quantum dynamical equations. However, as we explain above, in
the polymer picture of quantum mechanics the Hamiltonian operator
$H_{\hbar,\lambda}$ gets an additional parameter $\lambda$, which is
rooted in the ideas of minimum length. Now, note what happens if one
deals with the classical limit $\hbar \rightarrow0$ of this theory?
It is clear that although the parameter $\hbar$ will disappear from
the resulting classical Hamiltonian, this Hamiltonian is still
labeled by the parameter $\lambda$. Therefore, we are facing with a
classical theory that is not the one from which we begin and
described by the Hamiltonian $H$, but a classical theory with a
$\lambda$-dependent Hamiltonian $H_{\lambda}$. Of course, if all
things are well arranged, we expect to recover the initial classical
theory in the limit $\lambda \rightarrow 0$. It is believed that the
solutions of such an (lets call them) effective classical theories
can exhibit some important features of the corresponding dynamical
system which may be related to quantum effects and more importantly,
these effects can be achieved without quantization of the system. An
excellent explanation of this process can be found in \cite{CPR}.

In this paper, after a brief review of a flat FRW cosmology in the
presence of a massless scalar field, we introduce processes by means
of which this setup can be polymerized. We show that this issue may
be done by two equivalent methods. By one of them, to achieve a
$\lambda$-deformed Hamiltonian, we define a transformation in
phase-space which transforms the usual phase-space variables to
their polymer counterparts and by another, we find a non-canonical
coordinate transformation such that in terms of its new variables
the Hamiltonian retains its functional form but the corresponding
symplectic structure will be deformed. We then apply these two
methods to the above mentioned scalar field cosmological model to
arrive at the polymeric counterpart of the equations of motion. It
is shown that although these two methods of polymerization give
different equations of motion, the resulting scale factor and scalar
field, as the solution of the equations of motion, have the same
form independent of from which method they are obtained. We finally
summarize the work by a discussion around the possible relation
between the classical polymerization and quantum effects.

\section{Scalar field cosmology: a brief review}
In this section we consider the simplest dynamical model of
space-time, compatible with the assumption of a flat, homogeneous
and isotropic space on large scales. The geometry of such a
space-time is described by the Friedmann-Robertson-Walker (FRW)
metric with zero curvature index
\begin{equation}\label{A}
ds^2=-N^2(t)dt^2+a^2(t)(dx^2+dy^2+dz^2),
\end{equation}where $N(t)$ is the lapse function and $a(t)$ denotes
the scale factor which measures the expansion or contraction of the
universe in terms of the time variable $t$. By definition of the
cosmological proper time $\tau$ via $d\tau=Ndt$, the function $N(t)$
may be absorbed in time coordinate. Therefore, the above metric has
only one dynamical variable $a(t)$. If we also consider a free
scalar field $\phi(t)$ minimally coupled to the gravity, the
dynamics of the total cosmological setting will be
\begin{equation}\label{B}
{\cal S}=\int d^4x \sqrt{-g}\left(\frac{1}{16\pi
G}R-\frac{1}{2}g^{\mu \nu}\partial_{\mu}\phi
\partial_{\nu}\phi\right),
\end{equation}in which we have used the notation $g$, $G$ and $R$
for the determinant of the metric tensor $g_{\mu\nu}$, Newton
gravitational constant and the Ricci scalar respectively. The
Einstein field equations for the gravitational field and also the
scalar field Klein-Gordon equation can be obtained by variation of
the above action with respect to $g_{\mu \nu}$ and $\phi$
respectively. However, in order to pass to the polymerized
counterpart of the theory in the next section and compare the
results, we prefer to obtain the dynamics from the Hamiltonian
equations. The best way to do this issue is to follow the ADM
formalism and express the action in terms of the minisuperspace
variables $(a,\phi)$. By this method the Lagrangian (and hence the
Hamiltonian) of the model (see relations (\ref{C}) and (\ref{E})
below) is very similar to the one of a point particle moving in a
two dimensional space with coordinates $(a,\phi)$\footnote{In
general, as we will see in the next section, the polymerization
process is based on the Hamiltonian formalism of the classical or
quantum mechanics. Therefore, although the usual classical dynamics
can be viewed by other (equivalent) alternative formulations, this
is not the case when we are going to polymerize the system since
here only the Hamiltonian formalism works well. In this sense use of
the minisuperspace picture to construct the Hamiltonian for
description of the model is quite reasonable.}. So, by substituting
the Ricci scalar associated with the metric (\ref{A}), that is
\[R=6\left(\frac{\ddot{a}}{N^2a}+\frac{\dot{a}^2}{N^2
a^2}-\frac{\dot{N}\dot{a}}{a N^3}\right),\]and integration over
spatial dimensions, we obtain a point-like Lagrangian as
\begin{equation}\label{C}
L(a,\phi,\dot{a},\dot{\phi})=-\frac{3}{8\pi
G}a\dot{a}^2+\frac{1}{2}a^3\dot{\phi}^2,
\end{equation}in which we have set $N=1$ so that the time parameter
$t$ is the usual cosmic time. The momenta conjugate to the variables
$(a,\phi)$ can be calculated from their standard definition as
\begin{equation}\label{D}
p_a=\frac{\partial L}{\partial \dot{a}}=-\frac{3}{4\pi
G}a\dot{a},\hspace{5mm}p_{\phi}=\frac{\partial L}{\partial
\dot{\phi}}=a^3\dot{\phi},
\end{equation}from which by the usual canonical analysis one derives
the Hamiltonian
\begin{equation}\label{E}
H=-\frac{2\pi G}{3}\frac{p_a^2}{a}+\frac{p_{\phi}^2}{2a^3},
\end{equation}which is constrained to vanish due to gauge freedom of
the action. Now, we can write the classical dynamics from the
Hamiltonian equations, that is
\begin{eqnarray}\label{F}
\left\{
\begin{array}{ll}
\dot{a}=\{a, H\}=-\frac{4\pi G}{3}\frac{p_a}{a},\\\\
\dot{p_a}=\{p_a, H\}=-\frac{2\pi G}{3}\frac{p_a^2}{a^2}
+\frac{3p_{\phi}^2}{2a^4},\\\\
\dot{\phi}=\{\phi, H\}=\frac{p_{\phi}}{a^3},\\\\
\dot{p_{\phi}}=\{p_{\phi}, H\}=0.
\end{array}
\right.
\end{eqnarray}Since for a given cosmological setting, any solution
of the Einstein field
equations should satisfy the Hamiltonian constraint $H=0$, we may
consider this condition as the first integral of the above field
equations. So, with the help of (\ref{E}) the Friedmann equation
takes the form
\begin{equation}\label{G}
a^4\dot{a}^2-\alpha^2=0,
\end{equation}where $\alpha^2=4\pi G p_0^2/3$ and we take
$p_{\phi}=p_0=\mbox{const.}$ from the last equation of (\ref{F}).
This equation can easily be integrated to obtain the scale factor
as
\begin{equation}\label{H}
a_{\pm}(t)=(\pm 3\alpha t-t_0)^{1/3},
\end{equation}where $t_0$ is an integration constant. Assuming a positive value
for $t_0$, the condition $a(t)\geq 0$  would indicate that the
expressions of $a_{+}(t)$ and $a_{-}(t)$ are valid for $t\geq
\tau_0$ and $t\leq -\tau_0$  respectively, where
$\tau_0=t_0/3\alpha$. Therefore, by using of the third equation of
the system (\ref{F}), we get the complete set of the solution as
\begin{eqnarray}\label{I}
\left\{
\begin{array}{ll}
a_{-}(t)=(-3\alpha t-t_0)^{1/3},\\\\
\phi_{-}(t)=-\frac{p_0}{3\alpha}\ln(-3\alpha t-t_0),
\end{array}
\right.
\end{eqnarray}
for $t\leq -\tau_0$ and

\begin{eqnarray}\label{J}
\left\{
\begin{array}{ll}
a_{+}(t)=(3\alpha t-t_0)^{1/3},\\\\
\phi_{+}(t)=\frac{p_0}{3\alpha}\ln(3\alpha t-t_0),
\end{array}
\right.
\end{eqnarray}for $t\geq \tau_0$. These equations show that the
dynamical behavior of the universe with the scale factor $a_{+}(t)$
begins with a big-bang singularity at $t=\tau_0$ and then follows an
expansion phase at late times of cosmic evolution. For a universe
with the scale factor $a_{-}(t)$, on the other hand, the behavior is
opposite. The universe decreases its size from large values of scale
factor at $t=-\infty$ and ends its evolution at $t=-\tau_0$ with a
zero size singularity. In figure 1 we have shown the above mentioned
behaviors for these two sets of solution. It should be noted that
although we know that the universe is now in an expansion phase, the
contracting solution of the Einstein field equations are also
mathematically acceptable. However, the expanding and contracting
phases are disconnected from each other and should not consider as
simultaneously solutions of the system of field equations, i.e., the
expansionary phase will not occur after a contraction phase or vice
versa. In the next section we will see how this picture may be
modified if one enters the issues arising from an effective theory
into the problem at hand.

Before going to the next section to see how the above picture of the
classical model may be modified by means of the polymerization
mechanism, we would like to take a look at the Lagrangian equations
of motion correspond to the model. From (\ref{C}) the Euler-Lagrange
equations read as
\begin{eqnarray}\label{JJ}
\left\{
\begin{array}{ll}
\frac{d}{dt}\frac{\partial L}{\partial \dot{a}}-\frac{\partial L}{
\partial a}=0\Rightarrow \dot{a}^2+2a\ddot{a}+4\pi G a^2 \dot{\phi}^2=0,\\\\
\frac{d}{dt}\frac{\partial L}{\partial \dot{\phi}}-\frac{\partial
L}{\partial \phi}=0\Rightarrow \frac{d}{dt}(a^3 \dot{\phi})=0.
\end{array}
\right.
\end{eqnarray}Combination of the above equations yields
\begin{equation}\label{JJJ}
a^4\dot{a}^2+2a^5\ddot{a}+3\alpha^2=0.
\end{equation}It is easy too check that our previously set of
solutions (\ref{I}) and (\ref{J}) satisfy the Lagrangian equation of
motion (\ref{JJJ}). However, with $a_0$ and $t_0$ being some
integration constants and $\tau^2=9\alpha^2/4a_0^6$, this equation
also admits the following solutions
\begin{eqnarray}\label{JJJ1}
\left\{
\begin{array}{ll}
a(t)=a_0\left[(t-t_0)^2-\tau^2\right],\\\\
\phi(t)=\phi_0 \tanh^{-1}\frac{t-\tau}{t_0},
\end{array}
\right.
\end{eqnarray}which seems to be another set of solutions for our
problem at hand. But we should take care of that this solution does
not satisfy the Hamiltonian constraint $H=0$, and in this sense
cannot be considered as physical solution. Therefore, by the
Lagrangian formalism we are led again to the same solutions as we
have already obtained by the Hamiltonian equations of motion.

\section{Classical polymerization}
\subsection{Polymeric transformation}
We saw in the previous section how our classical model suffers from
the presence of a past or future big-bang like singularity, which
shows that some consequences may be physically unacceptable. It is
believed that dealing with such singularities would be in the
development of a quantum theory of gravity, according to which the
space-time, when it is considered in the high energy limit, has a
discrete structure due to the existence of natural cutoffs such as
the minimal measurable length. In Schr\"{o}dinger picture of quantum
mechanics, one usually feels free to work in the alternative
position or momentum space representations. However, in the presence
of the quantum gravitational effects the space-time manifold may
take a fuzzy structure so that the well-defined Schr\"{o}dinger
representations are no longer applicable \cite{GUP1}-\cite{Hossen}.
One of the recently proposed quantum frameworks which considers a
discrete nature for space in its formalism is the so-called polymer
representation of quantum mechanics \cite{Corichi1,Ashtekar}.
Instead of the Hilbert space ${\mathcal H}=L^2(R,dq)$, in which the
Schr\"{o}dinger quantum mechanics is formulated (here $dq$ is the
Lebesgue measure on the real line $R$), it is shown that the
appropriate Hilbert space to formulate the polymer quantum mechanics
is ${\mathcal H}_{\rm poly}=L^2(R_{_d},d\mu_{_d})$, where
$d\mu_{_d}$ is the Haar measure, and $R_{_d}$ denotes the real line
but now endowed with a discrete topology. The extra structure in
polymer picture is properly described by a dimension-full parameter
$\lambda$ such that the standard Schr\"{o}dinger representation will
be recovered in the continuum limit $\lambda\rightarrow\,0$
\cite{Corichi1}. Evidently, the classical limit of the polymer
representation $\hbar\rightarrow\,0$, does not yield to the
classical theory from which one has started but to an effective
$\lambda$-dependent classical theory which may be interpreted as a
classical discrete theory. Such an effective theory can also be
extracted directly from the standard classical theory (without any
attribution to the polymer quantum picture) by using of the Weyl
operator \cite{CPR}. The process is known as {\it polymerization}
with which we will deal in the rest of this paper.

In polymer representation of quantum mechanics, the position space
(with coordinate $q$) is assumed to be discrete with discreteness
parameter $\lambda$ and consequently the associated momentum
operator $\hat{p}$, that would be a generator of the displacement,
does not exist \cite{Ashtekar}. However, the Weyl exponential
operator (shift operator) correspond to the discrete translation
along $q$ is well defined and effectively plays the role of momentum
for the system under consideration \cite{Corichi1}. Taking this fact
into account, one can utilize the Weyl operator to find an effective
momentum in the semiclassical regime. Therefore, the derivative of
the state $f(q)$ with respect to the discrete position $q$ can be
approximated by means of the Weyl operator as \cite{CPR}

\begin{eqnarray}\label{FWD}
\partial_{q}f(q)\approx\frac{1}{2\lambda}[f(q+\lambda)-f(q-
\lambda)]\hspace{2cm}\nonumber\\=\frac{1}{2\lambda}\Big(
\widehat{e^{ip\lambda}}-\widehat{e^{-ip\lambda}}\Big)\,f(q)=
\frac{i}{\lambda}\widehat{\sin(\lambda p)}\,f(q),
\end{eqnarray}
and similarly the second derivative approximation gives
\begin{eqnarray}\label{SWD}
\partial_{q}^2f(q)\approx\frac{1}{\lambda^2}[f(q+\lambda)-2
f(q)+f(q-\lambda)]\hspace{1cm}\nonumber\\=\frac{2}{\lambda^2}
(\widehat{\cos(\lambda p)}-1)\,f(q).\hspace{2cm}
\end{eqnarray}Inspired by the above approximations, the
polymerization process is defined for the finite values of the
parameter $\lambda$ as
\begin{eqnarray}\label{Polymerization}
\hat{p}\rightarrow\,\frac{1}{\lambda}\widehat{\sin(\lambda p)},
\hspace{1cm}\hat{p}^2\rightarrow\,\frac{2}{\lambda^2}(1-
\widehat{\cos(\lambda p)}).
\end{eqnarray}This replacement suggests the idea that a
classical theory may be obtained via this process that is
dubbed usually as {\it Polymerization} in literature
\cite{Corichi1,CPR}
\begin{eqnarray}\label{PT}
q\rightarrow q,\hspace{1.5cm}p\rightarrow\frac{ \sin(\lambda
p)}{\lambda},\hspace{5mm}p^2\rightarrow
\frac{2}{\lambda^2}\left[1-\cos(\lambda p)\right],
\end{eqnarray}
and in a same manner one can find polymer transformation of the
higher powers of momentum $p$. In this sense, by a classical {\it
polymerized} system, we mean a system that the transformation
(\ref{PT}) is applied to its Hamiltonian. The first consequence of
the polymerization (\ref{PT}) is that the momentum is periodic and
its range should be bounded as
$p\in[-\frac{\pi}{\lambda},+\frac{\pi}{ \lambda})$. In the limit
$\lambda\rightarrow\,0$, one recovers the usual range for the
canonical momentum $p\in(-\infty,+\infty)$. Therefore, the
polymerized momentum is compactified and topology of the momentum
sector of the phase space is $S^1$ rather than the usual $R$
\cite{NaturalCutoff}.

\subsection{Polymeric symplectic structure}
Now, let us take a look at the Hamiltonian formalism in the context
of symplectic geometry. Given a configuration space ${\mathcal{Q}}$
of a dynamical system, the corresponding cotangent bundle
$T^{\ast}{\mathcal{Q}}$ (phase space) is naturally a symplectic
manifold endowed with a symplectic structure $\omega$ which, in
turn, is a closed non-degenerate $2$-form. In a local coordinate
$(q^1,...,q^n,p_1,...,p_n)$ on $T^{\ast}{\mathcal{Q}}$ the
symplectic structure takes the canonical form

\begin{equation}\label{canonicalstructure}
\omega=\sum_{i=1}^{n} dq^i\wedge dp_i\,,
\end{equation}where $(q^i,p_i)$ are called canonical variables.
The evolution of the system is given by

\begin{equation}\label{dy-eq}
i_{_{X_H}}\omega=dH.
\end{equation}Substituting of ${X_H}=\sum_{i=1}^{n}\big(\frac{{dq}^i}{
dt}\frac{\partial}{\partial q^i}+\frac{{dp}_i}{dt}\frac{\partial}{
\partial p_i})$ into the equation (\ref{dy-eq}) with canonical
structure (\ref{canonicalstructure}), the standard Hamilton's
equations for $(q^i(t),p_j(t))$ are emerged that are the integral
curves of $X_H$. For the minisuperspace
$(q^i,p_j)=(a,\phi,p_a,p_{\phi})$ we have considered in section 2,
these are indeed the set of equations (\ref{F}).

The triplet $(H,{\omega},X_H)$ constitutes a Hamiltonian system. In
order to study more complicated physical cases such as systems with
a noncommutative or polymeric structure, the corresponding
Hamiltonian system may includes some extra deformation parameters.
Here, we are going to claim that there are two distinct but
equivalent methods to enter the mentioned extra parameters into the
scenario:

$\bullet$ Approach I: In the first method we modify the Hamiltonian
to get a deformed Hamiltonian $H_{\lambda}$ where $\lambda$ is the
deformation parameter. In this method one does not change the
symplectic structure, so that the equations of motion can be
constructed as the usual case but now with the new Hamiltonian
$H_{\lambda}$. In the polymeric systems, for instance, this method
will be done by applying the polymerization transformation
(\ref{PT}) on the Hamiltonian. In this sense, in a two dimensional
phase space a typical Hamiltonian $H(q,p)=p^2/2m+U(q)$ takes the
form
\begin{equation}\label{PHamiltonian}
H_{_{\lambda}}=\frac{1}{m\lambda^2}\left[1-\cos(\lambda
p)\right]+U(q),
\end{equation}
with the use of which and the canonical symplectic structure
$\omega=dq\wedge dp$ in relation (\ref{dy-eq}) one is led to the
Hamilton's equations of motion $\dot{q}=\sin(\lambda p)/m\lambda$
and $\dot{p}=-\partial U/\partial q$.

$\bullet$ Approach II: In the second method, we seek a new set of
variables in terms of which the Hamiltonian takes its non-deformed
form. What is modified in this method is the symplectic structure
$\omega$ according to which the deformation parameter shows itself
in the equations of motion. To clarify this method in the polymer
framework, let us consider again the above two dimensional case and
apply the following  non-canonical transformation on its polymeric
phase space $\Gamma_{\lambda}$
\begin{equation}\label{NC-Transformation}
(q,p)\rightarrow\,(q',p')=\Big(q,\frac{2}{\lambda}\sin(\frac{
\lambda p}{2})\Big),
\end{equation}under the act of which the polymeric Hamiltonian
(\ref{PHamiltonian}) becomes $H=\frac{p'^2}{2m}+U(q')$ and the
corresponding symplectic 2-form will be
\begin{equation}\label{NSTF}
\omega_{_{\lambda}}=\frac{dq'\wedge\,dp'}{\sqrt{1-(\lambda
p'/2)^2}}\,.
\end{equation}
Since the canonical momentum $p$ varies in
a bounded domain, according to transformation
(\ref{NC-Transformation}) the range of the new momentum $p'$ should
be also bounded as $[-\frac{2}{\lambda}, +\frac{2}{\lambda})$. Also,
the new 2-form symplectic structure $\omega_{_{\lambda}}$ does not
have the canonical form and therefore the variables $(q',p')$ should
be considered as a pair of non-canonical variables. By means of this
set-up equation (\ref{dy-eq}) gives the associated Hamiltonian
vector field $X_H$ whose integral curves may be written as
$\dot{q'}=\frac{p'}{m}\,\sqrt{1-(\lambda p'/2)^2}$ and
$\dot{p'}=-\frac{\partial U}{\partial q'}\,\sqrt{1-(\lambda
p'/2)^2}$.

With a straightforward calculation based on the transformation
(\ref{NC-Transformation}) it is easy to show that the resulting
dynamics is the same regardless of whether it is obtained from the
first or the second approach. This means that by dealing either with
$(\omega_{\lambda},H)$ or with $(\omega,H_{\lambda})$ one leads to
the same $X_H$. This issue may be understood from the fact that
equation (\ref{dy-eq}) which defines $X_H$ is written in a
coordinate independent manner. Therefore, although $X_H$ has
different components when it is represented in terms of different
coordinates $(q,p)$ or $(q',p')$ that are related to each other by
(\ref{NC-Transformation}), as a geometrical object, $X_H$ has an
unique character independent of these coordinates. In the next
section we will apply the above mentioned classical polymerization
formalism to the scalar field cosmological model described in
section 2.

\section{Scalar field cosmology: polymeric dynamics}
In this section let us examine how the scalar field cosmological
setting in section 2 may change with polymeric considerations. As we
saw in section 2 this model has a four dimensional phase space whose
coordinates are $(a,\phi,p_a,p_{\phi})$ and its dynamics is given by
Hamiltonian (\ref{E}). It is important to note that we only
polymerize the geometrical part of the minisuperspace, that is
$(a,p_a)$, while the matter part $(\phi,p_{\phi})$ does not
contribute to our polymerization process \cite{CPR}. Therefore,
polymerization of this system according to the approach (I), will
lead us the effective Hamiltonian
\begin{equation}\label{A1}
H_{\lambda}=-\frac{4\pi G}{3\lambda^2 a}\left[1-\cos(\lambda
p_a)\right]+\frac{p_{\phi}^2}{2a^3}=-\frac{8\pi G}{3\lambda^2
a}\sin^2(\frac{\lambda p_a}{2})+\frac{p_{\phi}^2}{2a^3},
\end{equation}and the canonical symplectic structure

\begin{equation}\label{canonicalministructure}
\omega=da\wedge dp_a+d\phi\wedge dp_\phi.
\end{equation}Now, equation (\ref{dy-eq}) with

\begin{equation}\label{x}
X_H=\frac{da}{dt}\frac{\partial}{\partial a}+\frac{
{dp_a}}{dt}\frac{\partial}{\partial p_a}+\frac{{d\phi}}{dt}\frac{
\partial}{\partial \phi}+\frac{{dp_\phi}}{dt}\frac{\partial}{
\partial p_\phi},
\end{equation}yields the following equations of motion

\begin{eqnarray}\label{A2}
\left\{
\begin{array}{ll}
\dot{a}=-\frac{4\pi G}{3\lambda a}\sin(\lambda p_a),\\\\
\dot{p_a}=-\frac{8\pi G}{3\lambda^2 a^2}\sin^2(\frac{
\lambda p_a}{2})+\frac{3p_{\phi}^2}{2a^4},\\\\
\dot{\phi}=\frac{p_{\phi}}{a^3},\\\\
\dot{p_{\phi}}=0.
\end{array}
\right.
\end{eqnarray}On the other hand if we go ahead through the approach
(II), our starting point will be the non-deformed Hamiltonian
(\ref{E}) but now with the polymeric structure

\begin{equation}\label{minincstructure}
\omega_{_{\lambda}}=\frac{da\wedge\,dp_a}{\sqrt{1-( \lambda
p_a/2)^2}}+d\phi\wedge dp_\phi,
\end{equation}
which is obtained in the light of the deformed symplectic structure
(\ref{NSTF}). Clearly the matter part of the symplectic 2-form is
remained unchanged since the polymerization only acts on the
geometrical part of the minisuperspace. Substituting the Hamiltonian
(\ref{E}) and polymeric symplectic structure (\ref{NSTF}) together
with the Hamiltonian vector field (\ref{x}) into the relation
(\ref{dy-eq}) gives the following equation of motion for the triplet
$(H,\omega_{\lambda},X_H)$
\begin{eqnarray}\label{A2-1}
\left\{
\begin{array}{ll}
\dot{a}=-\frac{4\pi G}{3}\frac{p_a}{a}\sqrt{
1-(\frac{\lambda p_a}{2})^2},\\\\
\dot{p_a}=-\frac{2\pi G}{3}(\frac{p_a}{
a})^2+\frac{3}{2}\frac{p_\phi^2}{a^4},\\\\
\dot{\phi}=\frac{p_{\phi}}{a^3},\\\\
\dot{p_{\phi}}=0.
\end{array}
\right.
\end{eqnarray}
From the first equation of this system we have
\begin{equation}\label{A3}
\dot{a}^2=\frac{16 \pi^2 G^2 p_a^2}{9a^2}\Big(1-(\frac{
\lambda p_a}{2})^2\Big),
\end{equation}which upon substitution the expression
$p_a^2=\frac{3p_{\phi}^2}{4\pi Ga^2}$, from the constraint
equation $H=0$ one gets
\begin{equation}\label{A4}
\dot{a}^2=\frac{4\pi G p_{\phi}^2}{3a^4}\Big(1-\frac{3\lambda^2
p_{\phi}^2}{16\pi G a^2}\Big).
\end{equation}It is easy to show that this equation can also be extracted from the system (\ref{A2})
with the help of the constraint equation $H_{\lambda}=0$. Taking
into account from the last equation of (\ref{A2-1}) that
$p_{\phi}=p_0=\mbox{cons.}$, the above equation will be casted into
the form
\begin{equation}\label{A5}
a^3\dot{a}=\pm\alpha \sqrt{a^2-\beta^2},
\end{equation}where $\alpha^2=4\pi G p_0^2/3$ and $\beta^2=3\lambda^2
p_0^2/16\pi G$. This equation can be integrated to yield the
solution
\begin{eqnarray}\label{A6}
a_{\lambda}^{\pm}(t)=\bigg\{-\beta^2+\frac{2^{1/3}\beta^4}{\left[2\beta^6+
(\pm3 \alpha t-t_0)^2+ (\pm 3 \alpha t-t_0)\sqrt{4\beta^6+(\pm 3
\alpha t-t_0)^2}\right]^{1/3}}+ \nonumber\\
\frac{1}{2^{1/3}}\left[2\beta^6+(\pm 3 \alpha t-t_0)^2+ (\pm 3
\alpha t-t_0)\sqrt{4\beta^6+(\pm 3 \alpha t-t_0)^2}\right]^{1/3}
\bigg\}^{1/2},
\end{eqnarray}where $t_0$ is an integration constant. This expression provides a classical description of the scale factor as modified
by effective polymer dynamics. At the first glance it may seem that
it differs from the classical scale factor obtained in section 2 by
a complex relation. However, by a deeper look we realize that it
satisfies all of our expectations from a polymeric theory. First, it
is seen that in the limit $\lambda \rightarrow 0$ the above
expression takes the form of the ordinary case given in (\ref{H}).
Second, due to the presence of the parameter $\lambda$ in (\ref
{A6}), it escapes from the classical scale factor only in the region
where the scale factor tends to zero and for the large values of
scale factor coincides on it (see also figure 1 and discussion
below). To evaluate the dynamics of the scalar field, it is better
to expand $a_{\lambda}(t)$ as

\begin{equation}\label{A7}
a_{\lambda}^{\pm}(t)\sim (\pm3\alpha
t-t_0)^{1/3}-\frac{\beta^2}{2(\pm3\alpha t-t_0)^{1/3}}+O(\beta^3),
\end{equation}so that from the third equation of the system
(\ref{A2}) one obtains
\begin{equation}\label{A8}
\phi_{\lambda}^{\pm}(t)=\frac{p_0}{\alpha}\left[\pm\frac{1}{3}\ln\left(\pm
3\alpha t-t_0\right)-\frac{3}{4}\frac{\beta^2}{\left(\pm3\alpha
t-t_0\right)^{2/3}}+O(\beta^3)\right].
\end{equation}Now, let's see how the polymeric picture might lead to
a resolution of the primordial singularity appeared in the results
of section 2 and yields instead a bouncing connexion between
contracting and expanding phases. In figure 1 the scale factors are
plotted together with their non-deformed counterparts. In section 2
we have seen that the corresponding scalar field classical cosmology
admits two separate solutions, which are disconnected from each
other by a classically forbidden region. One of these solutions
represents a contracting universe ending in a singularity while
another describes an expanding universe which begins its evolution
with a big bang singularity. As this figure shows in the case where
the classical model is polymerized, the scale factor has a bouncing
behavior, i.e. the expansion phase in the cosmic evolution is
followed by a contraction phase. In this picture the classically
forbidden region is where the universe bounces from a contraction
epoch to a re-expansion era. It is clear that the reason for the
bouncing behavior in the vicinity of the classical singularity is
the existence of the $\lambda$-term in the polymeric model.
Therefore, if we consider the bouncing point as the minimum size of
the universe as suggested by quantum theories of cosmology, our
polymerization process support the idea that the polymeric
corrections to the classical cosmology are some signals from quantum
gravity.

\begin{figure}
\includegraphics[width=2.5in]{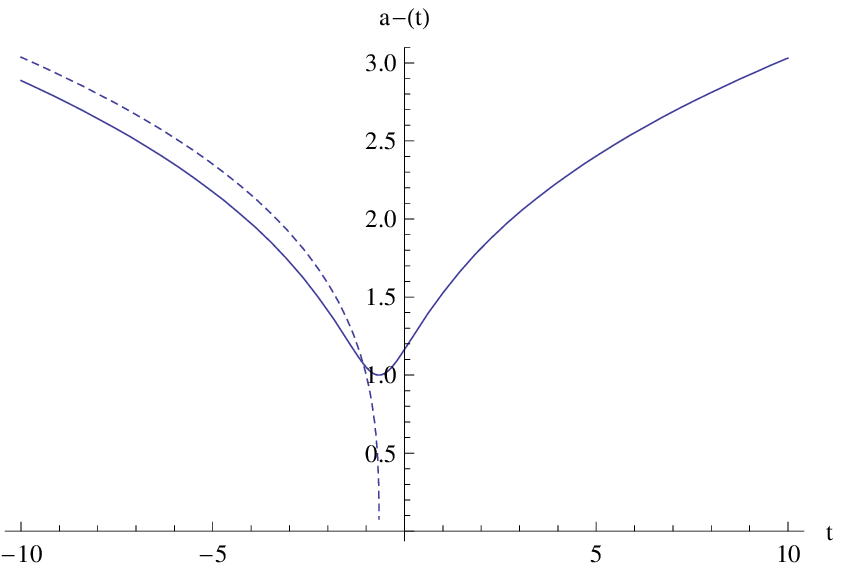}\hspace{3cm}\includegraphics[width=2.5in]{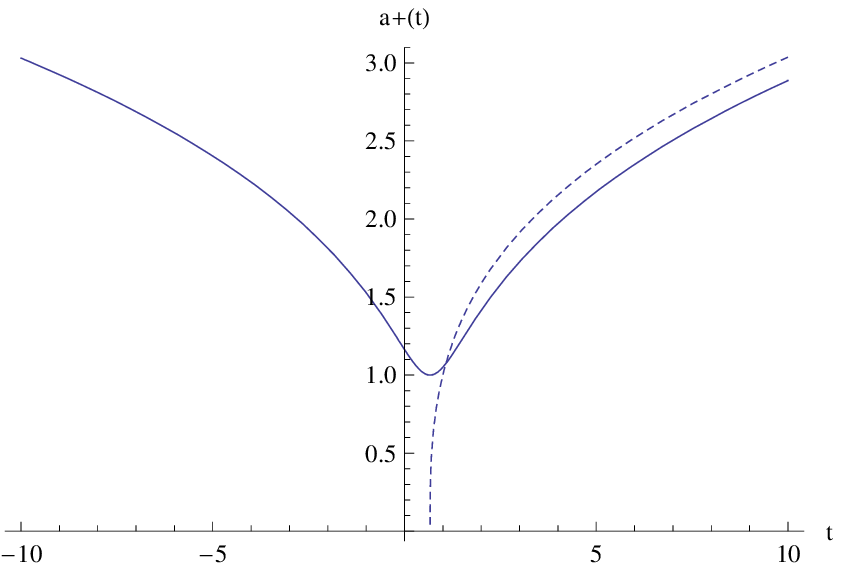}
\caption{Qualitative behavior of the scale factors versus time based
on relation \ref{H} (dashed lines) and \ref{A6} (solid lines).}
\end{figure}

\section{Summary}
In this letter we have studied the possibility of removing the big
bang singularity from a scalar field model of FRW cosmology by
introducing a classical process called polymerization. For this
purpose, we first reviewed a flat FRW geometry coupled to a massless
scalar field as a simple cosmological model which exhibits big bang
like singularity in its solutions. We have shown that this setup
admits two separate sets of solutions, while one of them is an
expanding universe begins its evolution from a big bang singularity,
the scale factor of another decreases its size from large values
until it eventually reaches a zero size singularity. These two sets
are separated from each other by a classically forbidden time
interval. Inspired by the polymer representation of quantum
mechanics, we then have dealt with a deformed classical theory in
which the momenta are transformed like their operator counterparts
in polymer quantum mechanics. We also presented an alternative
method to classically polymerized of the system, this time not by
changing the functional form of the momenta and therefore the
Hamiltonian, but by deforming the symplectic structure associated to
the corresponding Hamiltonian system. Finally, we applied classical
polymerization to the our minisuperspace model and solved the
resulting equations of motion once again. Interestingly, we found
that the scale factor displays a bouncing behavior, i.e., after a
period of contraction, an expansion era occurs. In the late time of
cosmic evolution the classical and the polymerized solutions will
coincide to each other. However, when the classical solutions
approaches their singularities the polymerized solutions get away
from them and bounce from a minimum value the size of which is
directly related to the polymeric deformation parameter. Bearing in
the mind that the prediction of such a bouncing behavior for the
scale factor is the main character of the quantum cosmological
models, we conclude that the classical polymeric structure which we
have constructed has a good correlation with quantum cosmology.

\end{document}